Metamateriały, konfigurowalne matryce antenowe i komunikacja holograficzna – wstępna analiza nowej koncepcji bezprzewodowej transmisji danych.

Dr hab. inż. Adrian Kliks, prof. uczelni

Instytut Radiokomunikacji, Politechnika Poznańska

e-mail: adrian.kliks@put.poznan.pl

tel. +48 61 665 3813

Streszczenie: W ostatnich kilku latach bardzo duże zainteresowanie wśród naukowców z całego świata zdobywa niezwykle oryginalna koncepcja komunikacji holograficznej. Specyfika tego podejścia z jednej strony niezwykle mocno odbiega od znanych i stosowanych obecnie rozwiązań, z drugiej stwarza bardzo duże możliwości rozwojowe w zakresie komunikacji bezprzewodowej. W artykule w sposób przeglądowy przedstawiono dwa rozwiązania technologiczne, które stały się przyczyną powstania idei komunikacji holograficznej. Jako pierwsze opisano możliwość wykorzystania tzw. meta materiałów dla celów bezprzewodowej transmisji danych, jako drugie natomiast - zastosowanie rekonfigurowalnych powierzchni antenowych. W ostatniej części przedstawiono z kolei założenia idei komunikacji holograficznej, w której znane z holografii optycznej zasady tworzenia obrazów zostały przeniesione do pasma radiowego i do pewnego stopnia uogólnione.

Abstract: In the last few years, a very original concept of holographic communication has gained a lot of interest among scientists from all over the world. The specificity of this approach, on the one hand, is very different from the known and currently used solutions, on the other hand, it creates great development opportunities in the field of wireless communication. The article provides an overview of two technological solutions that gave rise to the idea of holographic communication. First, the possibility of using the so-called meta materials for the purposes of wireless data transmission, and the second - the use of reconfigurable antenna surfaces. The last part presents the assumptions of the idea of holographic communication, in which the principles of creating images known from optical holography have been transferred to the radio band, and to some extend – generalized.

1. Wprowadzenie

Jednym z dominujących tematów w kontekście prowadzonych prac badawczych, wdrożeniowych i standaryzacyjnych w obszarze bezprzewodowych systemów telekomunikacyjnych było opracowanie rozwiązań dla tzw. piątej generacji (5G) sieci komórkowych [1]-[3]. Równocześnie w zakresie sieci lokalnych wiele prac podjętych było w celu upowszechnienia szóstej generacji (IEEE 802.11ax) oraz zdefiniowania nowych rozwiązań dla generacji siódmej (IEEE 802.11be) [4]-[8]. Równie dynamicznie rozwijają się inne rodzaje sieci, wystarczy wspomnieć tu o praktycznych implementacjach sieci Internetu Rzeczy [9]-[10], optycznej transmisji bezprzewodowej [11]-[13] czy komunikacji satelitarnej [14]-[15]. Należy zaznaczyć, że z perspektywy naukowej moment wdrażania i standaryzacji stanowi swojego rodzaju zakończenie pewnego etapu badań nad nowymi technikami czy nawet technologiami. Równolegle więc do wyżej wspomnianych działań społeczność naukowa pracuje nad rozwiązaniami, które być może znajdą zastosowanie w kolejnych (najbliższych, albo jeszcze późniejszych) generacjach systemów bezprzewodowych. Wystarczy wspomnieć, iż od kilku lat trwają intensywne dyskusje nad wymaganiami stawianymi szóstej generacji sieci komórkowych (6G), a także kolejnym generacjom sieci lokalnych [16]-[17]. W dyskusji tej wymienia się także różne przypadki zastosowań oraz rozwiązań technologicznych, które miałyby zagwarantować wypełnienie wspomnianych wymagań.

Wśród rozwiązań najczęściej pojawiających się w debacie naukowej należy wymienić m.in. powszechne użycie algorytmów sztucznej inteligencji (ang. *artificial intelligence*, AI), wykorzystanie możliwości wynikającej z komunikacji semantycznej (ang. *semantic communications*, SC) [18], zastosowanie programowalnych i otwartych sieci dostępu radiowego (Open Radio Access Networks, Open-RAN) [19] i WiFi (Open-WiFi) [20], rozwiązań kwantowych [21]-[22], czy też techniki rejestrów rozproszonych dla komunikacji bezprzewodowej [23]. Skupiając się na warstwie fizycznej systemu telekomunikacyjnego należałoby wymienić w szczególności modulację OTFS (ang. *Orthogonal Time Frequency Space*) [24], techniki zintegrowanej komunikacji i nasłuchiwania widma (ang. *Integrated Communications and Sensing, ICAS*) [25], pełnej, jednoczesnej komunikacji dupleksowej w tym samym paśmie (ang. *In-Band Full Duplex Communications*, IBFD) [25], wykorzystanie pasma bardzo wysokich częstotliwości w tym fragmentu pasma światła widzialnego (ang. *Visible Light Communications*) [11], integrację transmisji radiowej i optycznej (ang. Radio-Optical Communications) [27]-[30], czy wreszcie rekonfigurowalne matryce antenowe RMA lub inteligentne powierzchnie rekonfigurowalne (ang. *Reconfigurable Intelligent Surfaces*, RIS) [31]-[33].

Propozycje wykorzystania matryc antenowych składających się z bardzo dużej liczby elementów w telekomunikacji bezprzewodowej są różnorodne. Jednym z najprostszych i najczęściej spotykanych rozwiązań jest użycie RMA jako macierzy odbijających w sposób kontrolowany sygnał padający. Podejście takie pozwalałoby na np. takie przekierowanie sygnału padającego, aby został od odebrany w miejscu, w którym bez macierzy RMA byłby on niedostępny lub zbyt słaby. W takim ujęciu RMA jest elementem pasywnym (tzn. bez aktywnego wzmocnienia sygnału). W innym ujęciu RMA mogłoby zostać wykorzystane do wytłumienia sygnału padającego w taki sposób, aby zminimalizować moc sygnału interferującego w zadanej lokalizacji. Możliwości użycia matryc są jednak znacznie większe i zostaną one omówione w kolejnych rozdziałach artykułu. Na szczególną uwagę zasługuje użycie tzw. meta-materiałów (meta-elementów) w procesie tworzenia powierzchni RMA. Intensywny rozwój technik materiałowych połączony z rozwojem elektroniki pozwolił na uzyskanie bardzo ciekawych efektów z perspektywy propagacji sygnałów. Bardzo intensywne prace w obszarze RMA a także wyniki badań nad zastosowaniem metamateriałów dla komunikacji bezprzewodowej stały się przyczynkiem do powstania propozycji zupełnie nowego spojrzenia na transmisję sygnałów, nazywanej obecnie komunikacją holograficzną. Ponieważ nazwa może być myląca, należy podkreślić, że w swojej istocie komunikacja holograficzna nie jest to tylko zbiorem technik pozwalających na przesyłanie na odległość sygnału holograficznego. W niniejszym artykule skupiono się na przedstawieniu koncepcji komunikacji holograficznej, wskazując na jej postawy teoretyczne oraz wyzwania badawcze. W szczególny sposób uwzględniono możliwości zastosowania metamateriałów oraz wykorzystania matryc RMA w celu praktycznej realizacji koncepcji komunikacji holograficznej [34]-[36].

Artykuł składa się z trzech głównych części. W pierwszej z nich (rozdział 2) przedstawiono ideę meta-elementów i meta-materiałów, skupiając się na możliwości ich użycia w propagacji sygnału. W trzecim rozdziale skupiono się na analizie rozwiązań z zakresu powierzchni RMA, a w czwartym – przedstawiono koncepcję komunikacji holograficznej, dyskutując jej podstawy teoretyczne oraz wyzwania. Wnioski płynące z przedstawionej w artykule analizy zawarto w rozdziale piątym.

2. Metamateriały

Propagacja fal elektromagnetycznych w różnego rodzaju środowiskach jest w bezpośredni sposób zależna od parametrów opisujących właściwości tego środowiska, w szczególności wymienić tu należy przenikalność elektryczną i magnetyczną. Pierwsza z nich (ang. *permitivity*) oznaczana jest często poprzez $\varepsilon$ i jest wielkością łączącą indukcję pola elektrycznego $\vec{D}$ z natężeniem tego pola $\vec{E}$ $E$, $\vec{D} = \varepsilon \vec{E}$. Druga z nich (ang. *permeability*) oznaczana jako $\mu$ i oznaczającej zdolność danego środowiska do zmiany indukcji pola magnetycznego $\vec{B}$ w sytuacji zmiany natężenia tego pola $\vec{H}$, $\vec{B} = \mu \vec{H}$. Innymi słowy, wielkości te określają odpowiednio zdolność materiału do polaryzacji elektrycznej (do „tworzenia dipoli elektrycznych") oraz magnetycznej (do „tworzenia dipoli magnetycznych"). Z kolei wszystkie cztery wielkości charakteryzujące pole elektromagnetyczne są ze sobą połączone poprzez podstawowe dla telekomunikacji prawa Maxwella, opisujące zasady propagacji fal radiowych.

Występujące naturalnie i stosowane obecnie materiały można sklasyfikować m.in. z perspektywy posiadanych przez nie właściwości przenikalności magnetycznej i elektrycznej. Wielkości te definiują także szeroko stosowany współczynnik załamania zgodnie ze wzorem $n = \pm\sqrt{\varepsilon\mu}$ (lub częściej przedstawiany jako $n^2 = \varepsilon\mu$). Warto zauważyć, że współczynnik załamania może przyjmować wartości zespolone, a składnik urojony związany jest z absorpcją ośrodka. W sytuacji, kiedy wartości obu przenikalności są dodatnie, możemy mówić o materiałach występujących naturalnie w środowisku czy o materiałach o typowych właściwościach. W literaturze anglojęzycznej pojawia się termin DPS (*double positive*). W ośrodku opisanym jako DPS propagacja sygnału odbywa się zgodnie z typowymi właściwościami wykorzystywanymi w transmisji sygnałów. W przypadku kiedy albo przenikalność elektryczna albo magnetyczna jest ujemna, współczynnik załamania staje się wartością urojoną, a transmisja sygnałów jest w ogólności niemożliwa ze względu na występujące ekspotencjalne tłumienie ośrodka. Materiały takie nazywane są często z angielskiego jako ENG (*epsilon negative*) i MNG (*mu negative*). Ostatnią z rozważanych tutaj sytuacji jest wystąpienie jednocześnie ujemnych wartości obu podatności – zarówno magnetycznej, jak i elektrycznej. W takiej sytuacji propagacja sygnału znów staje się możliwa, a materiały te nazywane są jako DNG (*double negative*)[1]. Materiały o takich cechach są

---
[1] W zasadzie w anglojęzycznej literaturze przedmiotu można spotkać różne nazwy stosowane dla określenia materiałów o ujemnych wartościach przenikalności magnetycznej i elektrycznej. Są to m.in. NGV - negative group

możliwe do wyprodukowania przez człowieka, nie występują standardowo w środowisku [37], [38]. Zestawienie czterech klas materiałów przedstawiono na Rys. 1.

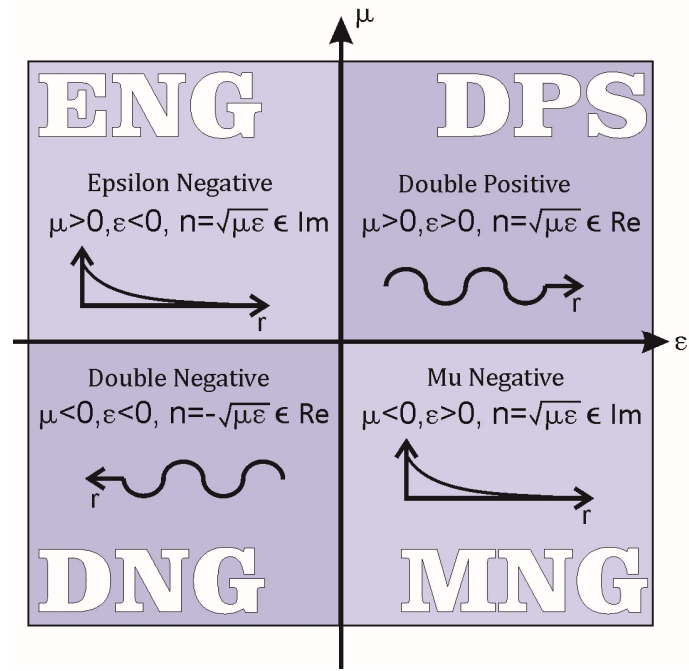

**Rys. 1. Podział materiałów ze względu na przyjmowane wartości przenikalności elektrycznej oraz magnetycznej**

W przypadku ośrodka określonego jako DPS, zarówno wektory opisujące natężenie pola elektrycznego $\vec{E}$ oraz magnetycznego $\vec{H}$ oraz wektor falowy $\vec{k}$ są prawoskrętne, innymi słowy wektor Poytinga i wektor falowy są skierowane tak samo. Sytuacja jest inna dla materiałów DNG, bowiem w tym przypadku wektor falowy jest przeciwnie skierowany do wektora Poytinga, a wektory $\vec{k}, \vec{E},$ oraz $\vec{H}$ są lewoskrętne. Sytuacja ta przekłada się na zupełnie nietypowe zachowanie się fali propagującej się w tym ośrodku. Warto tu wspomnieć o odwróconym zjawisku Dopplera (przy zbliżaniu się do źródła fali długość fali będzie ulegała zwiększeniu), negatywnym zjawisku załamania (soczewki wypukłe i wklęsłe działałyby w odwrotny sposób) czy też o fakcie, że prędkość grupowa fali będzie przeciwna do prędkości fazowej. Wspomniane własności materiałów DNG wskazują na możliwe ciekawe zastosowania w różnych dziedzinach życia, w tym w szczególności w telekomunikacji. Dla przykładu możliwość wpływania na zjawisko refrakcji mogłaby być z powodzeniem wykorzystania do kontrolowanego sterowania wiązką sygnału [40].

Jednakże możliwość wytworzenia materiałów DNG nie wyczerpuje samo w sobie możliwości oferowanych przez tzw. *metamateriały*. Słowo metamateriał wskazuje na materiał, którego cechy są „poza" tymi występującymi typowo w naturze. W ogólności więc, jako metamateriał można traktować każdą sztucznie wytworzoną i powtarzalną strukturę (materiał), a powtarzający się element w takiej strukturze nazywa się również także *meta-atomem*. Sztucznie wytworzony metamateriał nie musi być więc jednolity, może składać się z powtarzalnych struktur, które

---

velocity, NIM - negative index materials, DNM - double negative materials, LHM - left-handed materials. W języku polskim można spotkać nazwy np. materiały z ujemną refrakcją, materiały z ujemnym współczynnikiem załamania, podwójnie ujemne materiały, materiały z falą wsteczną czy wreszcie materiały z ujemną prędkością fazową.

traktowane razem będą składały się na powierzchnię metamateriału. Z perspektywy telekomunikacji warto zwrócić uwagę na ciekawy przykład budowy metamateriału składającego się z powtarzających się rezonatorów magnetycznych (*split-ring resonators*, SRR, prostych cewek odpowiedzialnych za reagowanie na zmiany pola magnetycznego) oraz małych dipoli (odpowiedzialnych za reagowanie na zmiany pola elektrycznego) [39]. Korzystając z tak zdefiniowanego *meta-atomu* udało się skonstruować trójwymiarowe struktury (*omni box*) pracujące na częstotliwościach rzędu GHz. Zwróćmy uwagę na fakt, że wytwarzanie metamateriałów w taki sposób otwiera nowe możliwości kontrolowania właściwości elektromagnetycznych tak wytworzonej struktury. W tym kontekście bardzo często wskazuje się, że odstęp pomiędzy sąsiadującymi elementami w strukturze powinien być z założenia mniejszy od czwartej części długości fali $\lambda/4$, jest to więc założenie inne niż stosowane powszechnie w kontekście technik wieloantenowych. W pracy [40] wskazano, że właściwości metamateriałowych struktur mogą być zmieniane na kilka sposobów. Pierwszy z nich to modyfikacja mechaniczna, gdzie zmienia się bezpośrednio odległość pomiędzy meta-atomami wpływając przez to na wypadkowe właściwości elektro-magnetyczne całej struktury. Jest to jednak metoda relatywnie kosztowna i dość złożona. Inne metoda modyfikacji polega na zmianie cech elektro-magnetycznych całego układu poprzez zmiany składowych elementów elektronicznych (wykorzystuje się tutaj np. diody PIN, waraktory, układy MEMS). Trzecia opcja zakłada zmianę właściwości struktury poprzez odpowiednie odziaływania temperaturowe, czwarta natomiast wykorzystuje różne możliwości oddziaływania chemicznego na użyte substancje składowe struktury (dość często wymienia się tutaj grafen).

Z kolei z perspektywy wytwarzania metamateriałów w pracy [40] wymieniono kilka metod wartych uwagi m.in. fotolitografia (stosowana dla jedno lub wielowarstwowych metamateriałów pracujących w zakresie długości fali od 30 nm do 3 mm), miękka fotolitografia, elektronolitografia (litografia wykorzystująca wiązkę elektronów) czy też SML (*shadow mask litography*), zwrócono także uwagę na techniki trójwymiarowe 3D.

Warto podkreślić, że rozwój metamateriałów stwarza szerokie możliwości aplikacyjne w różnych dziedzinach życia dzięki wykorzystaniu zarówno struktur określanych jako DNG, jak i tych składających się powtarzających się elementów składowych (*meta-atomów*). Wskazane wcześniej możliwości kontroli właściwości metamateriałów stały się inspiracją do powstania koncepcji materiałów programowalnych lub materiałów definiowanych programowo (*software defined materials* - SDM, *software programmable metasurfaces* - SPM) [41]-[43]. Przykładową strukturę sterowaną programowo zaprezentowano na Rys. XXXX. Analogicznie do idei radia czy sieci definiowanych programowo, zakłada się tutaj możliwość wpływania na właściwości powierzchni wykonanej z metamateriału z poziomu oprogramowania. Z kolei autorzy pracy [44] zaproponowali, aby najmniejszy powtarzalny element składowy (meta-atom) – nazywany super-komórką (*supercell*) - składał się z bardzo szerokiego wachlarza sub-elementów, gdzie każdy z nich jest odpowiedzialny za inne funkcje elementarne (np. pobieranie energii z otoczenia, dystrybucję energii w ramach elementu składowego, za komunikację w paśmie wysokich częstotliwości, komunikację w paśmie podczerwieni itp.).

Podsumowując powyższą jednak dość pobieżną analizę warto zwrócić uwagę na dynamikę rozwoju koncepcji metamateriałów, zwłaszcza z perspektywy ich zastosowania dla telekomunikacji. Pojawienie się materiałów z ujemną refrakcją z jednej strony oraz struktur fabrykowanych z wykorzystaniem powtarzalnych elementów składowych (meta-atomów) z drugiej pokazuje, że materiały te można potencjalnie wykorzystać do istotnego zmieniania warunków propagacyjnych nadawanych sygnałów telekomunikacyjnych. Materiały takie można użyć zarówno po stronie nadawczej i odbiorczej, jak i umieścić je w postaci matryc RMA bezpośrednio w środowisku propagacyjnym, wpływając na jego cechy. Należy jednak podkreślić, że wykorzystanie metamateriałów w przypadku RMA jest tylko jedną z możliwości ich zastosowania; matryce RMA mogą bowiem być rozmieszczone w środowisku jako fragment systemu telekomunikacyjnego niekoniecznie opierając się na metamateriałach (zwłaszcza tych określanych jako DNG).

3. Inteligentne matryce antenowe

W kolejnych generacjach systemów bezprzewodowych (komórkowych, lokalnych) coraz większą rolę odgrywały techniki wieloantenowe (Multiple-Input Multiple-Output), w których liczba elementów antenowych systematycznie wzrastała. Początkowo były to dwie lub kilka anten, później liczba ta wrosła do kilkunastu lub nawet kilkudziesięciu, aby ostatecznie mówić o technice *massive-MIMO*, gdzie liczba elementów antenowych jest co najmniej równa 128. W tym czasie równolegle rozpoczęła się dyskusja naukowa o możliwościach wykorzystania matryc antenowych RMA jako do pewnego stopnia niezależnych i autonomicznych elementów całego systemu telekomunikacyjnego. Wykorzystanie matryc antenowych RMA zazwyczaj łączna jest z koncepcją określaną z języka angielskiego jako inteligentne środowisko radiowe (*smart radio environment, SRE*). Autorzy pracy [45] zdefiniowali dwa bardzo istotne pytania, które w dokładny sposób odzwierciedlają założenia SRE:

- Jakie możliwości stworzyłaby możliwość monitorowania (tj. dokonywania obserwacji i raportowania pomiarów) środowiska bezprzewodowego poprzez rozmieszczenie urządzeń wykorzystujących do t ego celu tylko i wyłącznie sygnały już w tym środowisku istniejące, w dodatku bez konieczności użycia baterii?
- Co dałaby możliwość wpływania na środowisko propagacyjne (za pomocą kontrolera sterowanego zdalnie w sposób programowy i wyposażonego w możliwości predykcji) w celu np. zwiększenia przepływności bez zwiększania zużycia energii?

Pytania te mogą się w pierwszej chwili wydawać bardzo abstrakcyjne i dalekie od praktycznego zastosowania, jednak przy przyjęciu pewnych dodatkowych założeń pokazują ciekawe obszary badawcze z perspektywy propagacji fal radiowych, transmisji danych i modelownia systemów telekomunikacyjnych. Wzorując się na [45] na Rys. 2 przedstawiono klasyczne podejście do transmisji danych (gdzie zarówno po stronie odbiorczej i nadawczej podejmowane są działania w celu minimalizacji negatywnego wpływu kanału propagacyjnego), jak i podejście SRE (gdzie dodatkowo pojawia się opcja oddziaływania na środowisko propagacyjne).

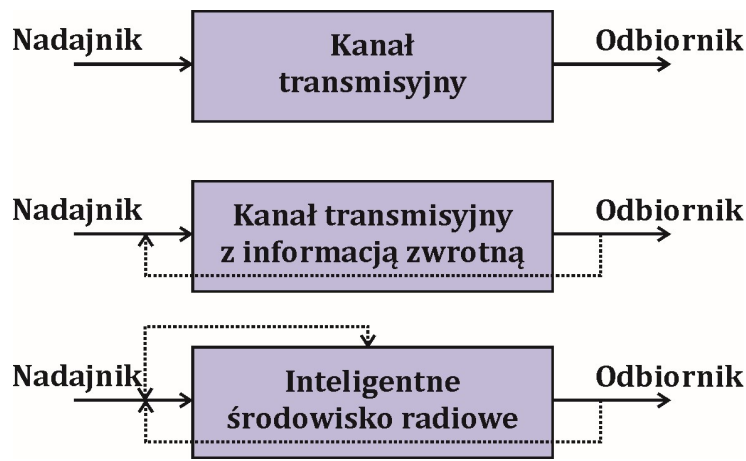

**Rys. 2. Ilustracja koncepcji inteligentnego środowiska radiowego (smart radio environment, SRE)**

Jedną z proponowanych technik wpływania na parametry środowiskowe w przypadku propagacji fal elektro-magnetycznych jest właśnie wykorzystanie możliwości stwarzanych przez meta-powierzchnie (czyli powierzchnie stworzone z wykorzystaniem meta-materiałów). W prostszym ujęciu, meta-powierzchnie mogą być zastąpione przez inteligentne powierzchnie rekonfigurowalne RIS lub też rekonfigurowalne matryce antenowe RMA. W tradycyjnym ujęciu RIS lub RMA składa się z wielu gęsto upakowanych i konfigurowalnych elementów antenowych, których odległość między nimi jest większa niż czwarta część lub połowa długości fali. Dodatkowo każdy z elementów antenowych jest sterowany za pomocą przesuwnika fazowego (ang. *phase shifter*) [37]. Dzięki takiemu podejściu RMA jest traktowany jako element pasywny (tzn. bez aktywnego wzmocnienia sygnału). Jak już wcześniej wspomniano, RIS czy RMA mogłoby zostać wykorzystane do wytłumienia sygnału padającego w celu minimalizacji interferencji w zadanej lokalizacji.

W literaturze naukowej można odnaleźć setki bardzo zaawansowanych publikacji związanych z zastosowaniem matrycRIS/RMA w telekomunikacji. Możliwości stwarzane przez powierzchnie RIS/RMA są na tyle obiecujące, że Europejski Instytut Norm Telekomunikacyjnych (European Telecommunications Standards Institute, ETSI) powołał specjalistyczną grupę roboczą zajmującą się tym zagadnieniem (ang. *Industry Specification Group (ISG) on Reconfigurable Intelligent Surfaces (RIS)*). Jako RIS rozumie się tutaj węzeł sieci bezprzewodowej wyposażony w specjalną powierzchnię zbudowaną z elementów antenowych lub metamateriałów, posiadającą specyficzne właściwości odbijające, refrakcyjne i absorpcyjne, które mogą być zmienione w zależności od panujących warunków środowiskowych. Grupa ISG-RIS zajmuje się wielorakimi zagadnieniami związanymi z użyciem RISów do komunikacji bezprzewodowej (w tym definicję przypadków użycia i scenariuszy rozmieszczenia, określenie wyzwań i możliwości technologicznych, zastosowanie do lokalizacji i tzw. sensingu, do zwiększenia bezpieczeństwa i prywatności w komunikacji, do minimalizacji ekspozycji na promieniowanie elektro-magnetyczne w wybranej lokalizacji itp.). Ponadto członkowie grupy starają się opracować scenariusz referencyjny przeznaczony dla kompletnego testowania systemów *end-to*-end, w których wykorzystuje się węzły RIS. W kwietniu i czerwcu 2023 r. w grupa w ramach ETSI opublikowała dwa raporty dotyczące możliwości zastosowania RISów [46]-[47]. Warto w tym miejscu zacytować, jak węzeł RIS jest interpretowany przez ISG RIS:

„*RIS can be implemented using mostly passive components without requiring high-cost active components such as power amplifiers, resulting in low implementation cost and energy consumption. This allows easy and flexible*

*deployment of RIS, with the possibility of RIS taking any shape and to be integrated onto objects (e.g. walls, buildings, lamp posts, etc.). RIS are supposed to run as nearly-passive devices and hence are unlikely to increase exposure to EMF, and in fact, they can potentially be used to reduce EM pollution in legacy deployments. These associated characteristics suggest RIS may be considered as a sustainable environmentally friendly technology solution. RIS may have different structures with considerations of cost, form factor, design and integration.*" [Cytat za [46]]

Macierze RMA można podzielić na aktywne i pasywne ze względu na technikę wytwarzania. W tym pierwszym przypadku mówimy o macierzach, w skład których wchodzą energochłonne elementy wysokoczęstotliwościowe oraz jednostki odpowiadające za przetwarzanie sygnałów. Z kolei powierzchnie pasywne działają jakby lustro metaliczne, które może być odpowiednio przeprogramowane w zależności od sposobu, w jaki ma zmienić właściwości fali padającej; typowo układy te nie potrzebują dodatkowego źródła zasilania. Warto zauważyć, że technika wytwarzania macierzy RMA nie wpływa na jej sposób pracy – macierz oznaczona jako aktywna może działać w trybie bez zapotrzebowania na dodatkową energię.

Skupiając się na możliwych trybach pracy, wskazano na następujące warianty [47]:

- odbijające powierzchnie RMA pozwalające na zmianę kąta odbicia (ang. *reflective surfaces*),
- refrakcyjne (załamujące) powierzchnie RMA pozwalające na zmianę kąta załamania (ang. *refractive surfaces*),
- absorbujące powierzchnie RMA używane w celu minimalizacji efektu rozpraszania (ang. *absorbing surfaces*)
- powierzchnie RMA posiadające obie powyższe własności, które mogą być wykorzystane jednocześnie (ang. *joint reflective and refractive surfaces*),
- powierzchnie RMA transmisyjne lub informacyjne, posiadające możliwość zakodowania danych oraz transmisji radiowej (ang. *transmitting and information surfaces*)
- powierzchnie odbijające i refrakcyjne pozwalające jednocześnie na modulację fali odbitej i załamanej sygnałem z danymi (*surface for ambient backscattering*)
- powierzchnie zwiększające rozpraszanie sygnałów w danym obszarze (ang, *surfaces for tuned randomness*)
- powierzchnie pozwalające na jednoczesne odbicie fali i wykrywanie obecności obiektów (ang. *communication and sensing surfaces*).

W literaturze odnaleźć można wiele ciekawych prac opisujących zarówno aspekty teoretyczne transmisji sygnałów w systemie wykorzystującym RMA, jak i praktyczne ich implementacje. Autorzy bardzo ciekawej pracy [48] zaproponowali wykorzystanie powierzchni IOS (ang. *Intelligent Omni Surfaces*), które posiadają zarówno możliwości odbijające i refrakcyjne zgodnie z Rys. 3.

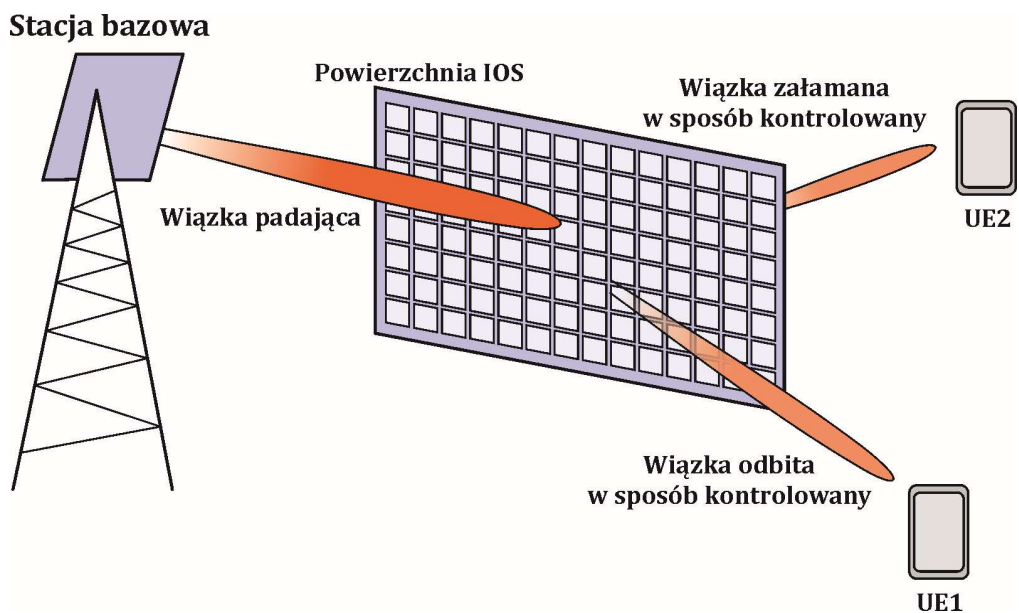

**Rys. 3. Powierzchnia IOS umożliwiająca zarówno kontrolowane odbicie jak i załamanie wiązki**

Działanie takich macierzy umożliwiałoby więc jednoczesne skierowanie fali odbitej oraz załamanej w pożądanych kierunkach. Właściwości odbijające zaproponowanej matrycy IOS zostały osiągnięte dzięki użyciu modułów składających się z diod PIN, promienników (*reflective patch*) oraz obszarów będących masą układu, a także metalowej osłony minimalizującej zjawisko ugięcia. W przypadku zjawiska refrakcji posłużono się dwoma takimi modułami (ze zmodyfikowanym promiennikiem) złączonymi ze sobą i połączonymi ze sobą szczeliną (*via hole*). Połączenie tych dwóch rozwiązań ze sobą jednocześnie dało Autorom możliwość osiągnięcia jednoczesnego kontrolowanego zjawiska odbicia i załamania. W artykule przedstawiono równoważny model elektryczny modułu IOS i opisano macierze przejść dla takiego modelu. Dodatkowo zaproponowano opis matematyczny modelu propagacyjnego, w którym założono obecność elementów IOS pozwalających na formowanie wiązek odbitej oraz załamanej. Co ciekawe, przedstawiono także wyniki eksperymentów pomiarowych z wykorzystaniem kontrolowanych macierzy IOS. Na szczególną uwagę zasługują opisane eksperymenty z użyciem macierzy IOS jako materiału na szyby w oknach umożliwiających sterowanie sygnałem padającym na tę „szybę".

Ciekawe wyniki eksperymentu pokazano w pracy [49], gdzie Autorzy za pomocą matrycy RMA umieszczonej na zewnątrz budynku byli w stanie istotnie zwiększyć moc sygnału obserwowanego w badanym obszarze. W szczególności rozważono sytuację, w której w wybranej lokalizacji za budynkiem jakość sygnału 5G mierzona za pomocą miary RSRP (ang. *Reference Signal Receive Power*) była niewystarczająca – właśnie z powodu dużego tłumienia spowodowanego przeszkodą (rogiem budynku i drzewami). Zmierzona wartość RSRP bez użycia powierzchni RMA była w badanym obszarze mniejsza niż -110 dBm lub nawet niż -120 dBm. Dzięki umieszczeniu na przenośnym maszcie matrycy RIS, w badanym obszarze wartość RSRP wzrosła do zakresu od -90 dBm do -105 dBm.

Możliwości oferowane przez matryc RMA są jednak bardzo szerokie. Bardzo interesujące podejście użycia matryc RMA pokazano w pracy [50], gdzie w konstrukcji elementu podstawowego matrycy (*meta atomu*) wykorzystano waraktory. Dzięki opisowi matematycznemu przedstawiono równoważną impedancję poszczególnego modułu, jak i powstałej płaszczyzny. Założono przy tym, że dzięki odpowiednim zmianom napięcia, impedancja poszczególnego układu podstawowego może zmieniać się w czasie (w pracy przyjęto, że może przyjąć jedną z czterech wartości). Innymi słowy impedancja ta stała się matematycznie funkcją czasu i można ją traktować jako sygnał informacyjny (bity informacyjne są odwzorowywane w odpowiednie wartości impedancji wypadkowej). W proponowanym rozwiązaniu Autorzy oświetlają tak stworzoną matrycę za pomocą wysokoczęstotliwościowej fali padającej, uzyskując przez to przeniesienie sygnału informacyjnego (impedancji matrycy zapisanej w funkcji czasu) do pasma wysokich częstotliwości bez wykorzystania mieszacza, filtrów oraz szerokopasmowego wzmacniacza mocy. Eksperyment sprzętowy wykonano z wykorzystaniem macierzy FPGA oraz płyt radia definiowanego programowo SDR (ang. *Software Defined Radio*). Schemat poglądowy proponowanego rozwiązania zilustrowano na

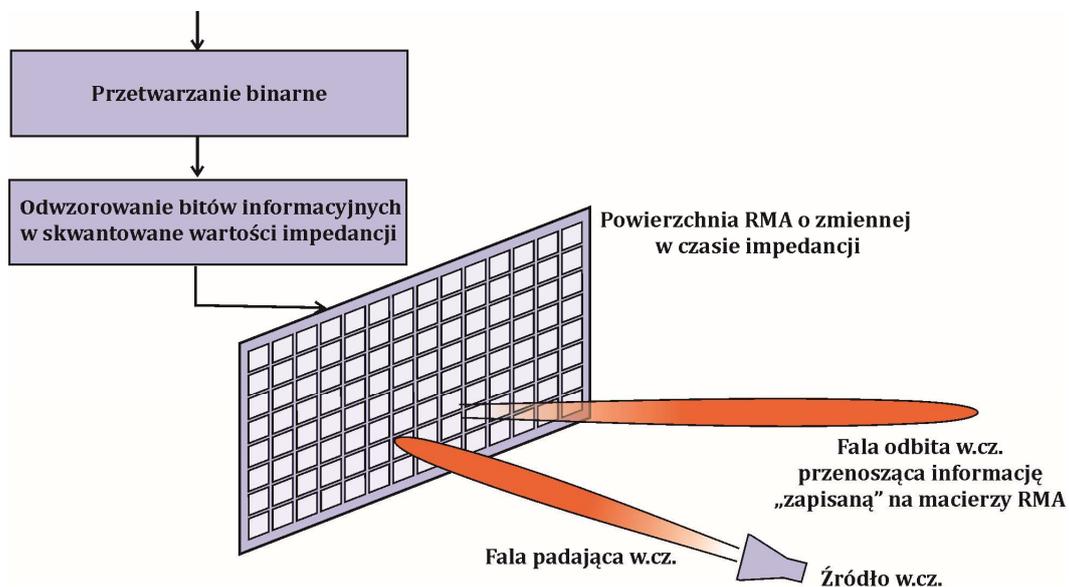

**Rys. 4. Ilustracja koncepcji przesyłania informacji zapisanej w postaci zmiennej impedancji płaszczyzny RMA**

Przytoczone powyżej przykłady pokazują bardzo szerokie spektrum zastosowania macierzy RMA do transmisji sygnałów. W tym ujęciu klasyczne rozumienie płaszczyzn RIS jako elementy tylko odbijające i pasywne okazuje się być tylko jednym z najprostszych wariantów ich wykorzystania w praktyce. Warto tutaj nadmienić, że literatura przedmiotu dostępna w bazach elektronicznych i dotycząca różnych aspektów macierzy RMA jest niezwykle bogata.

4. Komunikacja holograficzna

Przedstawiony ostatni przykład zastosowania macierzy RIS do transmisji sygnałów, w którym wyeliminowano z toru nadawczego układy mieszacza oraz filtry, wskazuje na nowy kierunek prac w obszarze telekomunikacji bezprzewodowej jakim jest tzw. komunikacja holograficzna. W tym miejscu należy podkreślić dwa spostrzeżenia. Pierwsze, komunikacja holograficzna w ujęciu przedstawianym w tym artykule nie jest transmisją sygnałów holograficznych (np. obrazów holograficznych, strumienia danych wykorzystywanych w okularach rozszerzonej rzeczywistości itd.) przez sieć bezprzewodową. Drugie, koncepcja komunikacji holograficznej jest na tyle nowym obszarem badawczym, że podczas studiów literaturowych można natrafić na zupełnie odmienne sugestie jej rozwoju proponowane przez różnych autorów, łącznie z różnicami dotyczącymi stosowanego słownictwa czy podstawowymi zjawiskami, które w przypadku komunikacji holograficznej powinno się rozważyć. W tym rozdziale przybliżymy podstawowe pojęcia i obserwacje związane z omawianym zagadnieniem.

W przypadku zastosowania macierzy RMA składającej się z wielu elementów podstawowych, zupełnie racjonalnym założeniem jest przyjęcie rozmiaru takiej macierzy $L$ liczonej w centymetrach albo nawet metrach. Dodatkowo można by założyć, że odległość pomiędzy nimi jest zdecydowanie mniejsza od połowy czy nawet czwartek części długości fali padającej, tak że pomiędzy elementami tej macierzy dochodzi do wzajemnych i istotnych oddziaływań elektro-magnetycznych. W konsekwencji efektywna apertura takiego układu będzie właśnie wyrażona także w centymetrach czy metrach [51]. Na Rys. 5 przedstawiono jedno z przykładowych zastosowań takiej macierzy jako elementu dużego elementu umieszczonego nad sceną lub estradą koncertową. Dodajmy, że zgodnie z analizą przedstawioną w poprzednich rozdziałach, macierze RMA mogą być wykonane z użyciem metamateriałów, których nawet obecnie grubość jest bardzo mała[2].

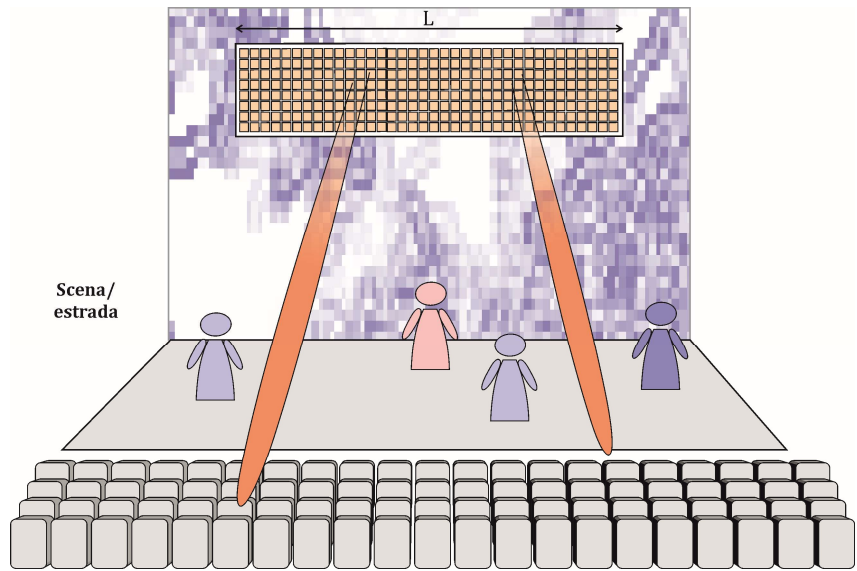

**Rys. 5. Przykład zastosowania płaszczyzny RMA o dużych wymiarach skutkujący możliwością wystąpienia transmisji danych w polu bliskim**

Zwróćmy uwagę, jakie w takiej sytuacji występują wartości granicy pomiędzy polem dalekim (ang. *far field*) a polem bliskim (ang. *near field*). Korzystając z prostych zależności [51], dla macierzy RMA o długości $L$=10 cm i dla częstotliwości około 10 GHz, granica między tymi strefami wynosi około kilku milimetrów. Jednak w przypadku zwiększenia rozmiaru macierzy do 100 cm, granica ta pojawia się w odległości prawie 100 m od środka macierzy. W praktyce oznacza to, że dla większych rozmiarów macierzy RMA (a takie są przecież rozważane), komunikacja będzie odbywała się bezpośrednio w polu bliskim. Na Rys. 5 zobrazowane jest to za pomocą dwóch wiązek, które w zasadzie mają charakter tylko ilustracyjny, bowiem transmisja w zależności od długości $L$ może odbywać się w polu bliskim. To zaś pociąga za sobą różnego rodzaju konsekwencje i zmusza do poszukiwania potencjalnie nowych rozwiązań z zakresu transmisji sygnałów uwzględniających szeroki wachlarz zjawisk pojawiających się w polu bliskim. W tym kontekście Autorzy wspomnianej pracy [51] proponują używanie terminu komunikacja holograficzna, aby podkreślić możliwości całościowego wykorzystania różnorodnych zjawisk pojawiających się w różnych odległościach od źródła (w polu bliskim i dalekim) i określenia limitów teoretycznych wynikających z własności propagacji fali elektro-magnetycznej w tych obszarach. Wskazują oni, że słowo holograficzny, wywodzące się z starogreckich słów *όλος* (*holos, all*), and *γραφή* (graphé, writing), oznacza literalnie "opisujący wszystko". Jest to jednak jedna z interpretacji dostępna w literaturze.

---

[2] Obrazowo można powiedzieć, że takie macierze RMA mogłyby być traktowane „naklejki" na ścianę, okno czy inne przedmioty, mogłyby być także traktowane jako rodzaj faktury ściany.

W bardzo obszernej pracy [52] (udostępnionej w trakcie tworzenia tego artykułu w repozytorium arXiv), Autorzy zaproponowali następującą definicję bezprzewodowej komunikacji holograficznej:

„*Holographic wireless communication—it is the physical process of realistically and completely restoring the three dimensional (3D) target scene transmitted by the transceiving ends with the help of new holographic antenna technology and wireless EM signal technology, and at the same time realize 3D remote dynamic interactions with people, objects and their surrounding environment.*" [cytat za [52]]

W definicji tej warto zwrócić uwagę na kilka aspektów, po pierwsze wspomniana jest przestrzeń trójwymiarowa przenosząca informację, po drugie – zakłada się wykorzystanie wszelkich zjawisk obecnych w propagacji fal elektro magnetycznych, po trzecie zaś – wspomniana jest koncepcja anten holograficznych. W pracy [53] zaproponowano używanie zwrotu Holographic MIMO Surfaces (HMIMOS), podkreślając fakt wykorzystania typowych zjawisk i rozwiązań stosowanych w tradycyjnej holografii optycznej do transmisji w paśmie fal radiowych. W tym przypadku macierz MIMO (także wykonana z metamateriałów, więc w istocie również powierzchnia RMA) pełni rolę powierzchni odpowiedzialnej za wytworzenie obrazu holograficznego w paśmie fal radiowych. Autorzy wzorując się na dwuetapowym procesie wytwarzania obrazów w holografii optycznej, proponują podobne podejście w odniesieniu do komunikacji bezprzewodowej. W szczególności w pierwszym etapie (nazywanym *holographic training*) zakłada się tutaj występowanie fali koherentnej, która rozdzielona na dwie składowe (falę padającą na obiekt oraz falę referencyjną) tworzą razem zestaw minimum i maksimów interferencji obserwowany i zapamiętany na płaszczyźnie HMIMOS. W drugiej fazie (nazwanej jako *holographic communications*) fala referencyjna odbita od płaszczyzny HMIMOS powoduje powstanie pożądanego obrazu holograficznego w odbiorniku. Zostało to zobrazowane na Rys. 6. Innymi słowy, w tym ujęciu komunikacja holograficzna stanowi swoiste przeniesienie rozwiązań znanych z holografii optycznej do obszaru fal radiowych.

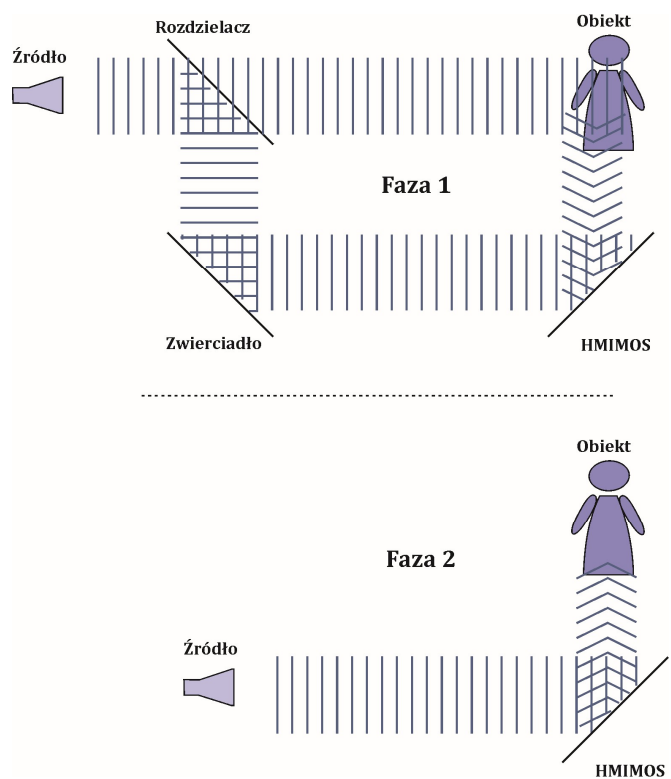

**Rys. 6. Ilustracja dwóch faz komunikacji holograficznej**

W konsekwencji takiego podejścia i z perspektywy analizy przesyłanej informacji, jak przedstawiono w [54], przestrzeń informacyjna staje się trójwymiarowa, bowiem odbiornik analizuje zachowania się wspomnianych ekstremów interferencji. Bardzo szczegółowy opis transmisji i odbiory takiego rodzaju sygnału przedstawiono m.in. w [52] oraz [55].

Wzorując się na [51] oraz [52] można zauważyć, że tradycyjne rozwiązania stosowane obecnie w telekomunikacji bezprzewodowej (dostosowane do propagacji fali w polu dalekim) zostają zastąpione rozwiązaniami typowymi dla pola bliskiego. W szczególności wykorzystując własności holografii opisane powyżej, rozważa się koncepcję, w której po stronie nadawczej stosuje się zbiór ortogonalnych funkcji dwuwymiarowych (opisujących np. rozkład prądów czy ogólniej pola elektromagnetycznego na powierzchni RMA, traktując tę powierzchnię jakby jako obraz), których rozkład na powierzchni RMA zależy od bitów informacyjnych. Obrazy takie są odtwarzane po stronie odbiorczej i analizowane z wykorzystaniem innego, dopasowanego zbioru ortogonalnych funkcji odbiorczych. W [52] z kolei zaproponowano wykorzystanie tensorów Green'a do przesyłania informacji w przypadku kanału LOS (Line of sight), zaś dla kanału NLOS (Non-LOS) - zastosowanie rozwinięcia sygnału odbieranego w dwuwymiarowy szereg Fouriera (odzwierciedlający falę płaską na powierzchni). Rozwiązanie dla kanału LOS/NLOS takie przedstawiono graficznie na Rys. 7.

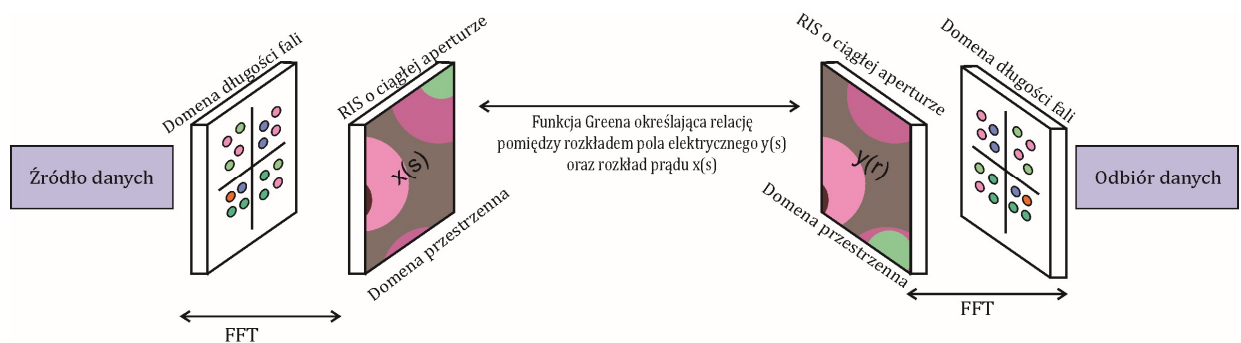

**Rys. 7. Przekazywanie danych w systemie z matrycami RIS (HMIMOS) dla kanału LOS oraz NLOS**

Bardzo ciekawym rozwiązaniem, inspirującym wielu badaczy, jest wykorzystanie orbitalnego moment pędu światła (ang. *orbital angular momentum,* OAM) znanego z sieci optycznych właśnie do komunikacji holograficznej z wykorzystaniem macierzy RMA [56].

Już pobieżna analiza rozwiązań stosowanych w komunikacji holograficznej wskazuje, że obserwujemy bardzo istotny rozwój nowej koncepcji bezprzewodowej transmisji sygnałów. W szczególności można zauważyć następujące różnice w komunikacji holograficznej względem rozwiązań tradycyjnych:

- w przypadku komunikacji holograficznej transmisja odbywać się może zarówno w polu dalekim, jak i bliskim, a ze względu na duże rozmiary powierzchni RMA w praktyce dominujący będą będzie transmisja w polu bliskim;
- pociąga to za sobą fakt, że czoło fali nie będzie mogło być opisywane i traktowane jako płaskie,
- z perspektywy macierzy RMA w komunikacji holograficznej zakłada się, że odległość pomiędzy elementami składowymi jest znacznie mniejsza od połowy długości fali; jest to zupełne przeciwieństwo typowych założeń stosowanych w technice wieloantenowej, gdzie minimalna odległość pomiędzy elementami aktywnymi macierzy MIMO powinna wynosić właśnie połowę długości fali. Ma to zapewnić brak wzajemnego oddziaływania elementów antenowych na siebie, a przez to możliwość traktowania kanałów propagacyjnych pomiędzy każdym elementem macierzy nadawczej a odbiornikiem jako niezależne;
- w tradycyjnym systemie wieloantenowym efektywna apertura antenowa jest dyskretna, w przypadku zaś macierzy HMIMOS można mówić o rozwiązaniach bliskich antenom o ciągłej aperturze (a jednocześnie pozwalających na uzyskanie wąskich wiązek propagacyjnych)
- w przypadku macierzy RMA (w wersji HMIMOS) liczba elementów składowych (*meta atomów*) jest z założenia bardzo duża, w efekcie czego – przy bardzo niewielkiej odległości między elementami – efektywna apertura jest także duża.

Wspomniane powyżej różnice pomiędzy systemami tradycyjnymi a tymi wykorzystującymi komunikację holograficzną wskazują, że potrzeba prowadzenia intensywnych prac naukowych dotyczących różnych ich aspektów. Wystarczy tutaj wskazać np. na konieczność opracowania efektywnej i niezawodnej metody odwzorowywania informacji użytkownika w przestrzeń holograficzną, gdzie informacja przechowywana jest w postaci ekstremów pojawiających się w wyniku oddziaływania fali referencyjnej. W literaturze przedmiotu można znaleźć propozycje zastosowania zaawansowanych przetworników radiowo-optycznych pozwalających na wykorzystanie cech światła widzialnego w komunikacji holograficznej. W innych wariantach obraz holograficzny jest wytwarzany jednak w paśmie fal radiowych. Niezależnie jednak od tego, bardzo ważnym problemem staje się zaproponowanie wydajnych metod estymacji i korekcji kanału bezprzewodowego na transmitowanych sygnał. Dodatkowo, współcześnie dominującym rozwiązaniem jest zapewnienie komunikacji do jednego ewentualnie niewielkiej liczby użytkowników; stąd istnieje konieczność dalszych badań w zakresie wielodostępu w systemach z komunikacją holograficzną. W istocie jednak potrzebne są także prace nad propozycją modelu propagacyjnego dla kanału bezprzewodowego, uwzględniające właściwości elektro-magnetyczne ośrodka względem propagującej się fali, a także nowe badania nad holograficzną teorią informacji, teorią kodowania czy synchronizacji. Warto tu podkreślić, że prace nad komunikacją holograficzną prowadzone są bardzo szeroko i niejednokrotnie proponowane rozwiązania wzbudzają wzmożoną dyskusję naukową. Jest to jednak obszar badawczy na tyle nowy a jednocześnie na tyle obiecujący, że autorzy wspomnianej już pracy [52] sugerują konieczność wypracowania nowej teorii transmisji danych, która będzie w sposób całościowy uwzględniała właściwości elektro-magnetyczne w systemie telekomunikacyjnym.

5. Podsumowanie

Z perspektywy komunikacji bezprzewodowej przedstawione w artykule trzy rozwiązania technologiczne mogą być rozważane zarówno niezależnie, jak i łącznie. Metamateriały, wytworzone z materiałów o ujemnym współczynniku załamania lub też z jednostek elementarnych (*meta atomów*), mogą zostać wykorzystanie w celu wpływania na właściwości propagacyjne środowiska. Mogą też stanowić podstawę do wytworzenia rekonfigurowalnych powierzchni RMA, które z kolei mogą także zostać zbudowane w tradycyjny sposób z typowych obecnie elementów antenowych stosowanych w rozwiązaniach MIMO. Powierzchnie RMA dostarczają same w sobie ogromnych możliwości aplikacyjnych, a ich właściwości i możliwości zastosowania stały się w ostatnich jednym z najintensywniej rozważanych tematów naukowych na całym świecie. Wreszcie przedstawiona relatywnie niedawno propozycja całościowego wykorzystania właściwości wszystkich zjawisk pojawiających się podczas transmisji bezprzewodowej sygnałów wydaje się kreować zupełnie nowe obszary badawcze. Komunikacja holograficzna jako koncepcja transmisji danych jest dopiero w początkowej fazie rozwoju, a wiele aspektów związanych z przekazywaniem informacji w ten sposób wymaga intensywnych prac, wiele jest także rozbieżności oraz wątpliwością z nią związanych. Wydaje się jednak, że pojawiające się tutaj możliwości są na tyle obiecujące, że coraz częściej komunikację holograficzną wymienia się jako jedną z technik wykorzystywanych w przyszłych systemach bezprzewodowych.

Podziękowanie

Literatura